# Ex-situ atomic force microscopy on the growth mode of $SrRuO_3$ epitaxial thin film

Bora Kim[1,#], Sang A Lee[2,#], Daehee Seol[1], Woo Seok Choi[2*] and Yunseok Kim[1*]

[1]*School of Advanced Materials Science and Engineering, Sungkyunkwan University (SKKU), Suwon, 16419, Republic of Korea*

[2]*Department of Physics, Sungkyunkwan University (SKKU), Suwon, 16419, Republic of Korea*

[#] B. K. and S. A L. contributed equally to this work.

[*] Address correspondence to choiws@skku.edu and yunseokkim@skku.edu

**Abstract**

The functional properties of devices based on perovskite oxides depend strongly on the growth modes that dramatically affect surface morphology and microstructure of the hetero-structured thin films. To achieve atomically flat surface morphology, which is usually a necessity for the high quality devices, understanding of the growth mechanism of heteroepitaxial thin film is indispensable. In this study, we explore heteroepitaxial growth kinetics of the $SrRuO_3$ using intermittent growth scheme of pulsed laser epitaxy and ex-situ atomic force microscopy. Two significant variations in surface roughness during deposition of the first unit cell layer were observed from atomic force microscopy indicating the possible formation of the half unit cell of the $SrRuO_3$ before the complete formation of the first unit cell. In addition, layer-by-layer growth mode dominated during the first two unit cell layer deposition of the $SrRuO_3$ thin film. Our observation provides underlying growth mechanism of the heteroepitaxial $SrRuO_3$ thin film on the $SrTiO_3$ substrate during the initial growth of the thin film.

**Introduction**

Heteroepitaxial perovskite oxide thin films have attracted a lot of scientific attention for their potential application for opto-electronic and energy devices. Intriguing and useful materials properties including superconductivity, 2-dimensional electron liquid (2DEL), multiferroicity, catalytic activity, and thermoelectricity have been tailored and realized.[1-3] It has frequently been shown that these physical and chemical properties depend strongly on surface morphology, atomic termination layer, and microstructure of the thin film, and in many cases, atomically flat surfaces and controlled substrate/film interface were a prerequisite for the intriguing and useful emergent phenomena.[2, 4-9] The epitaxial growth modes depend on several parameters such as the interface energy (surface adhesive force vs. adatom cohesive force), growth temperature and pressure, growth rate, step-terrace width (miscut angle) of the substrate, and the lattice mismatch between substrate and layer.[10-16] Commonly, epitaxial growth modes can be classified as layer-by-layer (LBL), step flow (SF), step bunching, and island growth.[10] In the LBL growth mode, new layers are nucleated only after completion of the layers below. This growth mode requires the low interface energy and small lattice mismatch and occurs when the average terrace width is larger than the diffusion length. The SF growth is characterized by the steady advancement of the step edges in the vicinal direction. The layers in the SF mode have relatively high crystalline perfection because defects are prevented due to coalescence. Usually, SF growth occurs when the average terrace width is smaller than the diffusion length of the adatoms. However, when the terrace width is too small, step bunching might occur by the crowding of the steps as the growth progresses.

Understanding the growth mechanisms of the epitaxial thin films and atomic scale control of the surface morphology and interface are of prime interest in studying the perovskite oxide heterostructures. In general, in-situ monitoring such as reflectivity high-energy electron diffraction (RHEED) and scanning tunneling microscopy (STM) is used to investigate the growth kinetics of the epitaxial thin films.[6, 17-19] More recently, growth monitoring of thickness and optical properties is also realized using in-situ spectroscopic ellipsometry.[20, 21] RHEED has been mostly used to observe the surface structure, thickness and control the growth rates in real time during a deposition.[17, 18] However, it is rather difficult to obtain actual surface morphology information such as clusters sizes, densities, and/or distributions. Also, RHEED intensity oscillation for the monitoring of the thickness is only possible for the LBL growth. While STM has the ability to observe the surface morphology, *e.g.*, recent STM study reported on the growth mode transition from a LBL to SF in perovskite oxide thin film growth,[19] it can only measure conducting samples for an extremely localized regions.[19, 22] As an alternative approach, atomic force microscopy (AFM) can be considered to investigate the growth kinetics for relatively broader regions. While it is technically rather challenging to monitor the growth in real-time, AFM enables the observation of the detailed surface morphology in a relatively wide region of the thin film sample. Furthermore, recently, it was reported that the variation of surface roughness could be analyzed using AFM,[23] indicating the usefulness of the technique in studying the growth mechanism and surface morphology.

SrRuO$_3$ (SRO) is one of the most widely used perovskite oxide electrode materials for oxide electronics. It is an itinerant ferromagnet ($T_C$ = ~160 K) based on the strong hybridization of Ru 4$d$ and O 2$p$ orbitals.[3] The structure of bulk SRO is orthorhombic with the space group of *Pbnm*, which can be considered as pseudocubic with the lattice constant of $a_{pc}$ = 3.926 Å. When grown on a SrTiO$_3$ (STO, $a$ = 3.905 Å) substrate, the lattice mismatch between STO substrate and SRO thin film is 0.54% (compressive strain). Because of the small lattice mismatch with STO substrate and good chemical stability in air, SRO is a suitable material for the study of the growth mechanism with AFM. It has been reported that SRO epitaxial thin film shows a change of growth mode during the initial deposition on STO substrates.[24, 25] Specifically, the growth mode changes from LBL to SF, accompanying the surface termination switching from $BO_2$ ($RuO_2$) to AO (SrO) which was confirmed by RHEED monitoring and STM.[19] Such transition in the growth mode and surface termination atomic layer was explained based on the reduction of surface migration barrier which is related to the chemical configuration of the interface.

In this study, we systematically studied the growth kinetics of SRO heteroepitaxial thin film grown on the $TiO_2$-terminated STO substrate using ex-situ AFM. Heteroepitaxial SRO thin films were intermittently grown on $TiO_2$-terminated (001) STO substrate using PLE. A variation of surface morphology during the fractional u.c. layer growth was investigated using ex-situ AFM.[26] The LBL growth mode occurs without growth mode transition during the first 2 u.c. layer deposition, indicating the consistent growth kinetics on the STO substrate.

**Experimental Procedure**

**Thin film fabrication and basic characterizations.**

High-quality epitaxial $SrRuO_3$ (SRO) thin film was grown on $TiO_2$-terminated (001) $SrTiO_3$ (STO) single crystalline substrates using pulsed laser epitaxy (PLE).[3, 27] The miscut angle of STO substrate was ~0.1°. A laser (248 nm; IPEX 864, Lightmachinery) fluence of 1.5 J/cm$^2$ and repetition rate of 2 Hz was used. The SRO thin film was grown stoichiometric condition, where is in oxygen partial pressure ($P(O_2)$) = $10^{-1}$ Torr at 700ºC. The rate of the temperature increase was 20ºC/min. For cooling, we turned off the halogen lamp heater immediately after the growth, to minimize surface diffusion of adatoms at elevated temperatures. The growth of epitaxial SRO thin film was confirmed using X-ray diffraction (XRD), atomic force microscopy (AFM), electric transport, and magnetic measurements. To monitor the growth kinetics using AFM, we employed an intermittent growth scheme realizing fractional u.c. layer deposition. After every 10 or 20 laser shots, the growth was ceased for the AFM investigation of the surface morphology ex-situ. We minimized the time of exposure to the atmosphere, to prevent surface contamination.[28]

**AFM characterization of the growth kinetics.**

The non-contact topography measurements were performed with a commercial AFM (XE7, Park Systems) using a non-contact AFM probe (NCHR, Nanoworld). The indentation was performed on the as-grown STO substrate before deposition of SRO thin film through a stiff diamond coated tip (CDT-NCHR, Nanoworld) to write a marker for measuring the same regions. The original AFM topography was measured as the higher image resolution (1024 × 1024 pixels) with 5 μm × 5 μm than general image resolution (256 × 256 pixels). The image processing was performed through commercial software (XEI, Park Systems) and MATLAB based homemade analysis program. The original images were cropped twice based on the indentation as the indicator (see Fig. S1) to clearly display terrace feature. We firstly cropped the original images from 5 × 5 μm$^2$ (1024 × 1024 pixels) to 4 × 4 μm$^2$ (819 × 819 pixels) and then, the image of 0.5 × 1 μm$^2$ (102 × 204 pixels) with five terraces of STO substrate was once more cropped from the firstly cropped images.

**Results and discussion**

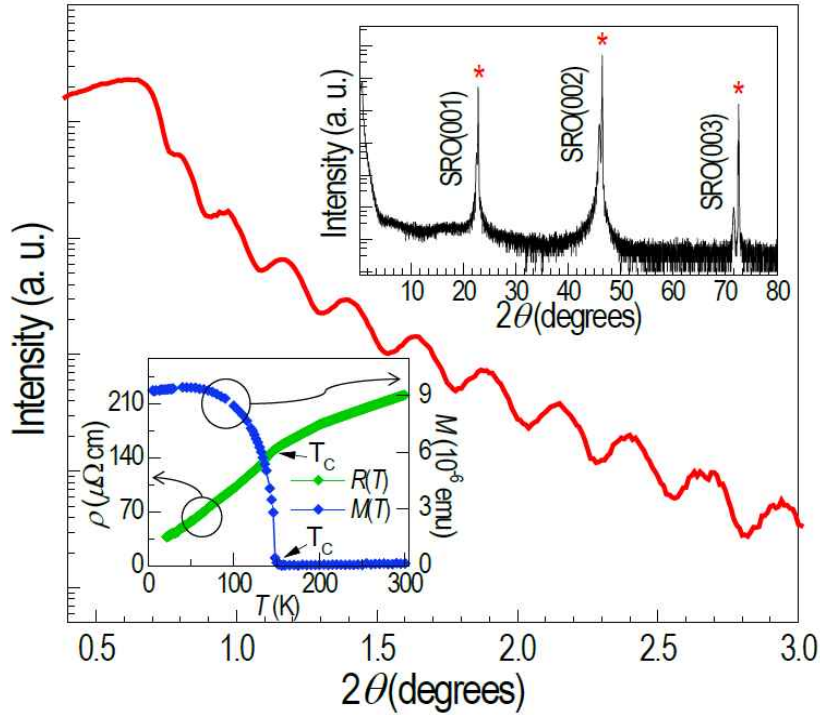

**Figure 1.** Stoichiometric SrRuO$_3$ epitaxial thin film evidenced by structural, electric, and magnetic characterization. The main figure shows x-ray reflectivity (XRR) result which indicates a ~30 nm-thick SrRuO$_3$ epitaxial thin film with smooth surface and interface. X-ray diffraction (XRD) $\theta$-$2\theta$ scan in the upper right inset shows that the single-crystalline epitaxial SrRuO$_3$ thin film is grown without any secondary phase. The bottom left inset shows typical resistivity and magnetization (at 100 Oe) curves as functions of temperature, clearly indicating the expected ferromagnetic transition at ~150 K for a stoichiometric SrRuO$_3$ thin film on a SrTiO$_3$ substrate.

We have fabricated heteroepitaxial stoichiometric SRO thin films on TiO$_2$-terminated STO (001) substrates using PLE. The x-ray reflectivity (XRR) result in Fig. 1 shows clear interference fringes indicating a ~30 nm-thick SRO epitaxial thin films with well-defined surface and interface. The x-ray diffraction (XRD) *θ-2θ* scan in the upper right inset of Fig. 1 shows epitaxial SRO thin film on STO substrate, without any secondary phases. The SRO thin films have good crystalline quality with a full-width-at-half-maximum (FWHM) value of < 0.02°. From XRD reciprocal space mapping (RSM) measurement, the SRO thin films are grown coherently on the STO substrate, with the same in-plane lattice constant (data not shown). The temperature dependent resistivity *R*(*T*) and magnetization *M*(*T*) results clearly show the ferromagnetic transition temperature at ~150 K, suggesting that we have a stoichiometric SRO epitaxial thin film on a STO substrate, in good agreement with other previous studies.[3, 29] From the total thickness of the thin film obtained from XRR and the number of laser shots fired, we estimated ~65.4 shots for the growth of 1 u.c. layer of SRO thin film. Based on this result, we employed an intermittent growth scheme and deposited fractional u.c. layer of SRO (10 to 20 laser shots) before the AFM investigations of the growth kinetics. In general, RHEED is used to investigate the initial stage nucleation and growth of the thin films. However, as mentioned previously, the SF growth nature of SRO prevents such studies using RHEED. Alternatively, we applied ex-situ AFM measurement to explore the growth kinetics of SRO, which provides direct information on surface morphology.

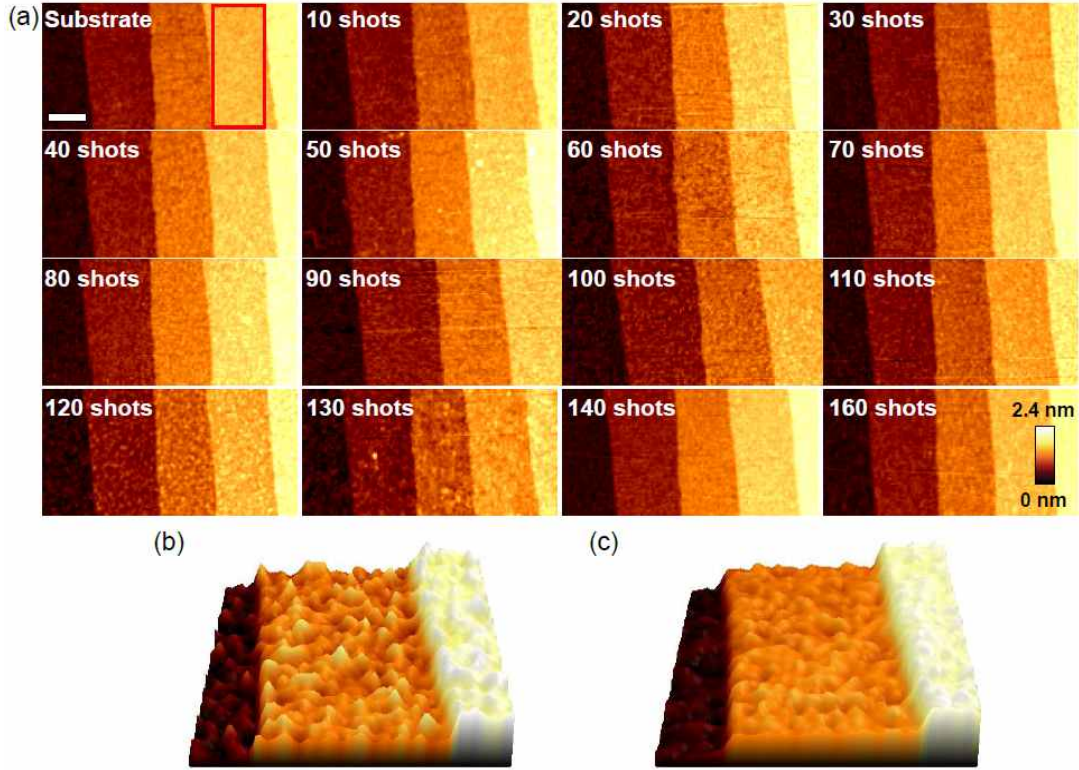

**Figure 2.** (a) Topography (102 × 204 pixels) of the SrRuO$_3$ on the SrTiO$_3$ substrate after each growth sequence with a laser interval of 10 or 20 shots. Note that these images were cropped from the original images of 1024 × 1024 pixels as shown in Fig. S1. Scale bar in Fig. (a) represents 150 nm. Topographical 3D images at (b) 120 and (c) 140 shots, respectively.

Figure 2 shows topography 2D and 3D images after deposition of subsequent fractional layers of SRO thin film. To investigate the initial growth kinetics of the epitaxial SRO thin film on the STO substrate, we observe the variation of the surface feature from STO substrate up to 160 laser shots with an interval of 10 or 20 shots. To monitor the same spot, we marked a square (dimension of 1 × 1 μm$^2$) on the bare STO substrate as the spatial reference, using a stiff diamond coated tip (nominal spring constant ~ 42 N/m) (see Fig. S1). The original images (1024 × 1024 pixels) were cropped as the images with five terraces (102 × 204 pixels) to clearly display the growth of SRO thin film on the STO substrate, as shown in Fig. 2 (a). The details related to the image processing can be found in Fig. S1. Note that a severely contaminated surface (exposed to air for more than 2 weeks) would result in an AFM topography image which is not possible to analyze (see Fig. S3). The AFM image of the contaminated surface indirectly suggests that our AFM measurements are relatively free from the contamination issue.

The variation of surface feature up to 70 shots, *i.e.*, the nominal formation of the first u.c. layer of SRO, is rather ambiguous. In contrast, during formation of the second u.c. layer (from 80 to 140 shots), the surface feature changes significantly. Distinguishable blobs are generated on top of the first u.c. layer up to 100 shots, and then surface relatively flattens out at 110 shots. These results indicate that the surface roughness is systematically changed during film deposition. Such behavior is clearly observed in the topographical 3D images of 120 and 140 shots, as shown in Figs. 2(b) and (c), respectively. The observed variation of the surface feature indicates that SRO layer is obviously deposited on top of the STO substrate.

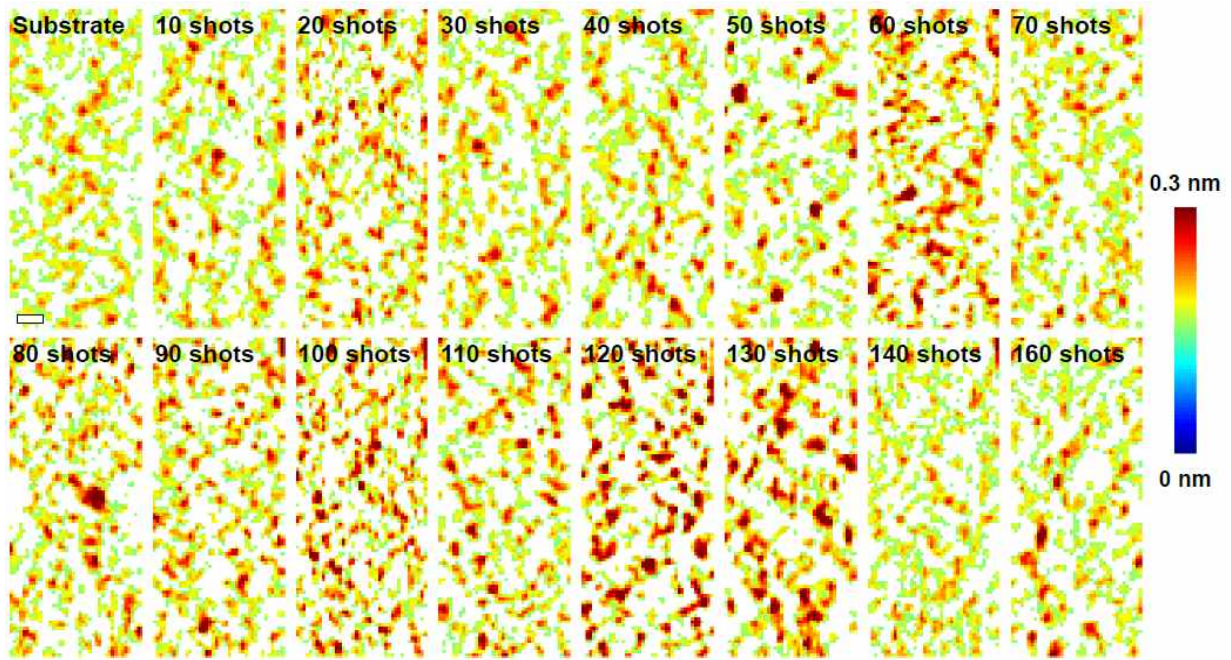

**Figure 3.** Filtered images from the original one terrace images, as shown in Fig. S2. The pixels of height above 0.15 nm are shown as color scale and other pixels are indicated as a white color at each shot. The scale bar is 40 nm.

For more detailed investigation of the surface feature, we implemented image filtering to the original topography images of one single terrace, as shown in Figure 3. The pixels of height above 0.15 nm are shown as color scale and other pixels are indicated as white color. These results directly provide the formation of the SRO layer through the variation of surface height, since deposited regions have relatively higher height than those regions without any deposition. While the diffusivity values of SRO adatoms reported in the literature are rather scattered, ranging from ~$7 \times 10^3$ to $1.1 \times 10^7$ nm$^2$/s [10, 12], the actual observation of the inhomogeneous growth suggest that the effect of surface diffusion is efficiently minimized during our experiment. The filtered topography shows similar tendency between the formation of the first (from 10 up to 70 shots) and second (form 80 up to 140 shots) u.c. layers, indicating that the analyses based on the topography images are well consistent with that estimated growth rate from XRR. While the contrast in the AFM topography images are very low in height, *i.e.*, 0.2~0.4 nm, and the general AFM measurement has an error range of < 0.4 nm, the formation of the SRO layer can be estimated from the systematic and periodic variation of the surface roughness, expected from other experimental observation (XRD). Another important observation is that the growth mode of epitaxial SRO thin film on STO substrate is LBL, at least up to 2 u.c. layer thickness. While SF growth is characterized by the movement of step edges during the growth, we could not observe any distinguishable changes of the step positions between the topography images of STO substrate and SRO thin films after 140 laser shots (Fig. S1). We note that previous research reported the growth mode transition from 2D LBL to SF without clearly identifying the exact point in time, and that 2D islands start to nucleate near the step-edge regions in the earliest stage growth.[24] However, our results show the nucleation is evenly distributed without favorable sites for nucleation.

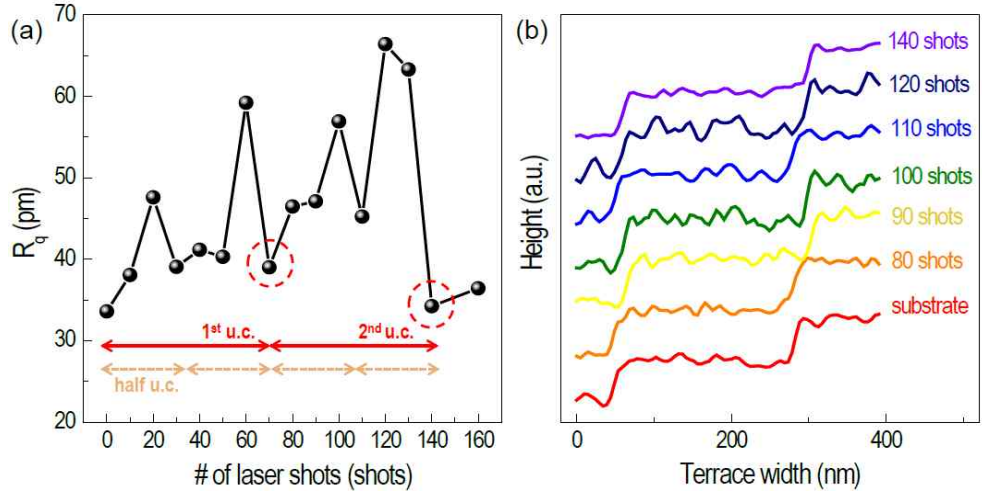

**Figure 4**. (a) Surface roughness and (b) area profiles of the SrRuO$_3$ thin film within the same terrace indicated by a red rectangle in Fig. 2, respectively. The variation of surface roughness shows the formation of 2 unit cell of SrRuO$_3$ on the SrTiO$_3$ substrate. Note that outskirts of one terrace are slightly included to visualize clearly terrace feature in area profiles.

In order to further investigate the growth kinetics of SRO epitaxial thin films on STO substrates, we analyzed surface roughness and area profiles. Figure 4(a) shows the root mean square roughness ($R_q$) from the STO substrate (0 shots) up to 160 laser shots of SRO deposition which were extracted from a single terrace indicated by a red rectangle in Fig. 2.[23] Two qualitatively similar cycles can be observed, indicating the formation of two u.c. layers. Within each u.c. layer growth sequence, two local maximum $R_q$ values are observed (20/60 and 100/130 shots), possibly suggesting the growth of half u.c. layer. The surface re-flattens out at 70 and 140 shots, respectively, for the first and second u.c. layers. Therefore, we can consider that the systematic change in the surface roughness originates the generation of the SRO blobs on the surface and then progressively flattens out as the density of generated blobs increases. The same tendency is also observed in the area profiles as shown in Fig. 4(b). When the film has minimum roughness, the step height converges within ~1 Å. On the other hand, when the film has maximum roughness, the step height of AFM images is about half u.c. (1.7 ~ 2.0 Å). This result further supports the formation of half u.c. layer during the growth of SRO film on STO substrate. In addition, during the initial growth, any delay in time for completing the u.c. layers is not observed, suggesting the absence of the surface termination switching.[12]

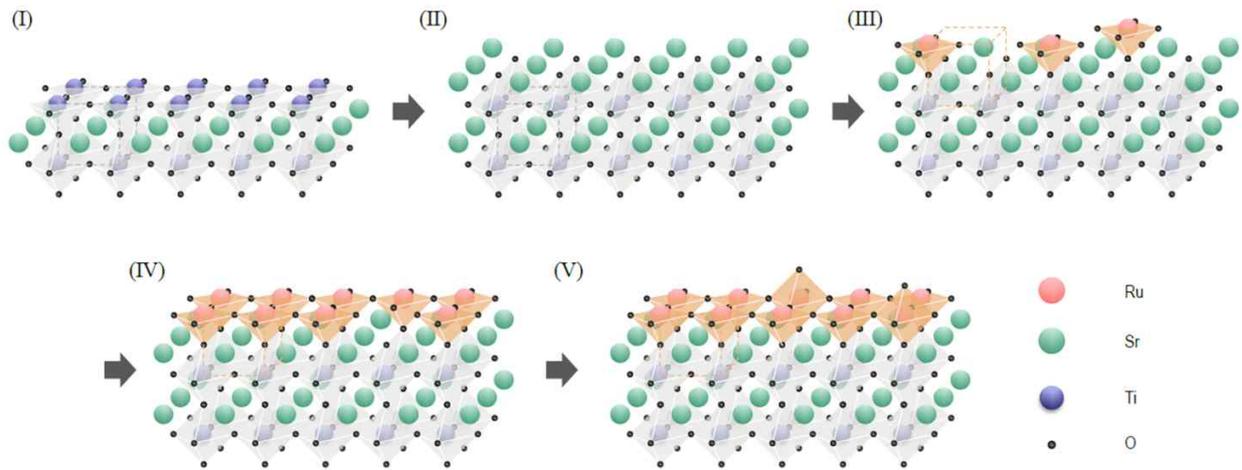

**Figure 5.** Schematic diagram of the growth mechanism of $SrRuO_3$ thin film on the $TiO_2$-terminated $SrTiO_3$ substrate.

On the basis of the AFM analyses, the growth mechanism of the SRO on the STO substrate can be understood. Figure 5 shows a schematic representation of the expected growth mechanism for the initial growth of the SRO thin film. The first and second u.c. layers show nearly the same tendency from the AFM results. The two peaks in the surface roughness for each u.c. layer growth implies that there are two steps during the formation of one u.c. layer. The (001)-oriented SRO within the pseudocubic notation consists of alternating atomic layers of SrO and $RuO_2$. In the first step, the $TiO_2$-terminated surface of STO preferentially serves as the platform for the formation of SrO layer (Fig. 5(I)). This leads to the change in the surface termination from $TiO_2$- to SrO-termination during the deposition of the half u.c.[12] The first increase of surface roughness should be due to the partial deposition of the SrO layer on top of the STO substrate, and the subsequent decrease of surface roughness in vicinity of 30 shots is related to the formation of SrO-terminated layer (Fig. 5(II)).[19, 30, 31] When the $TiO_2$ termination changes to SrO termination, another incomplete half u.c. ($RuO_2$) layer starts to be deposited (Fig. 5(III)). Finally, the formation of one u.c. of SRO can be observed (Fig. 5(IV)), further leading to second u.c. layer deposition. Within this growth scheme, LBL growth mode seems to dominate for the deposition of the first and second u.c. layer of SRO thin film on the STO substrate.

Our result provides important findings regarding the initial pulsed laser epitaxy of SRO thin film on STO substrate. First, the LBL growth is preferred for the initial SRO thin film deposition. The change in the growth mode to SF growth does not occur at least up to two u.c. SRO layers. This observation is consistent with previous reports, where the RHEED oscillation could be observed at least up to several u.c. layers.[12] STM study also did not show any signs of SF growth during the initial deposition of SRO on STO.[19] While it is possible that the intermittent growth scheme might modify the growth mode,[32] consistent observations from at least three different experimental techniques allow us to apply the ex-situ AFM technique to general PLE of transition metal oxides. Second, we postulate that the SRO layer can be deposited as half-a-unit-cell unit. Such information is impossible to obtain using RHEED, since RHEED is based on the diffraction technique which requires a whole u.c. layer for the detection of the thickness. The direct observation of the half u.c. layers based on AFM is consistent with the recent STM study, while Chang *et al*. suggested that the termination switching from $RuO_2$ to SrO also occurs, and the SrO capping layer may stabilize the $RuO_2$ layer.[19] The half u.c. layer growth obviously depends on different growth parameters, and can further be applied to the growth of transition metal oxide superlattices with the change of surface termination layers.

**Conclusion**

In conclusion, we investigated growth mechanism of the epitaxial $SrRuO_3$ thin film on the $SrTiO_3$ substrate using ex-situ AFM. Based on the known growth rate, we employed intermittent growth scheme of the $SrRuO_3$ thin film from the $SrTiO_3$ substrate up to 160 shots, equivalent to 2.3 unit cells. layers, with an interval of 10 and 20 shots for AFM investigation. It is observed that 2 unit cells layers are formed when 140 laser pulses are applied to the $SrRuO_3$ target, which is well consistent with XRR result. Furthermore, a variation of surface roughness reveals that half unit cell of $SrRuO_3$ layer might be firstly deposited and then, a unit cell of $SrRuO_3$ is completed. We confirm that the LBL growth mode occurs during deposited 2 unit cell on the $SrTiO_3$ substrate. Our observation provides the underlying growth mechanism of epitaxial $SrRuO_3$ thin film on the $SrTiO_3$ substrate and possibility for the control of sub unit cell growth.

**Acknowledgements**

This work was supported by "Human Resources Program in Energy Technology" of the Korea Institute of Energy Technology Evaluation and Planning (KETEP), granted financial resource from the Ministry of Trade, Industry & Energy, Republic of Korea. (No. 20174030201800). Also, this work was supported by Basic Science Research Programs through the National Research Founda

Supplementary Information

# Ex-situ atomic force microscopy on the growth mode of SrRuO$_3$ epitaxial thin film

Bora Kim[1,#], Sang A Lee[2,#], Daehee Seol[1], Woo Seok Choi[2*] and Yunseok Kim[1*]

[1]*School of Advanced Materials Science and Engineering, Sungkyunkwan University (SKKU), Suwon, 16419, Republic of Korea*

[2]*Department of Physics, Sungkyunkwan University (SKKU), Suwon, 16419, Republic of Korea*

[#] B. K. and S. A. L. contributed equally to this work.

* Address correspondence to choiws@skku.edu and yunseokkim@skku.edu

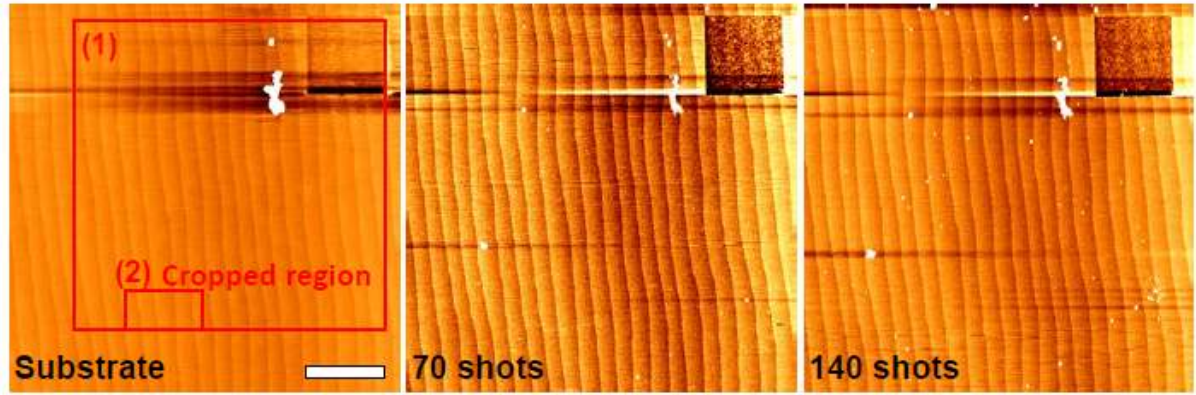

**Fig. S1.** AFM topography (1024 × 1024 pixels) of SRO on STO substrate as a sequence of a substrate, 70 and 140 shots. The indentation was performed through a stiff diamond coated tip at the top right corner to measure the same regions. Scale bar is 1 μm.

The indentation was performed on the STO substrate through stiff diamond coated tip to measure the same region and then, we cropped the original images twice based on indentation as shown in Fig. S1. We firstly cropped the original images from 5 × 5 μm$^2$ (1024 × 1024 pixels) to 4 × 4 μm$^2$ (819 × 819 pixels) and then, the image of 0.5 × 1 μm$^2$ (102 × 204 pixels) with five steps was cropped from the firstly cropped images.

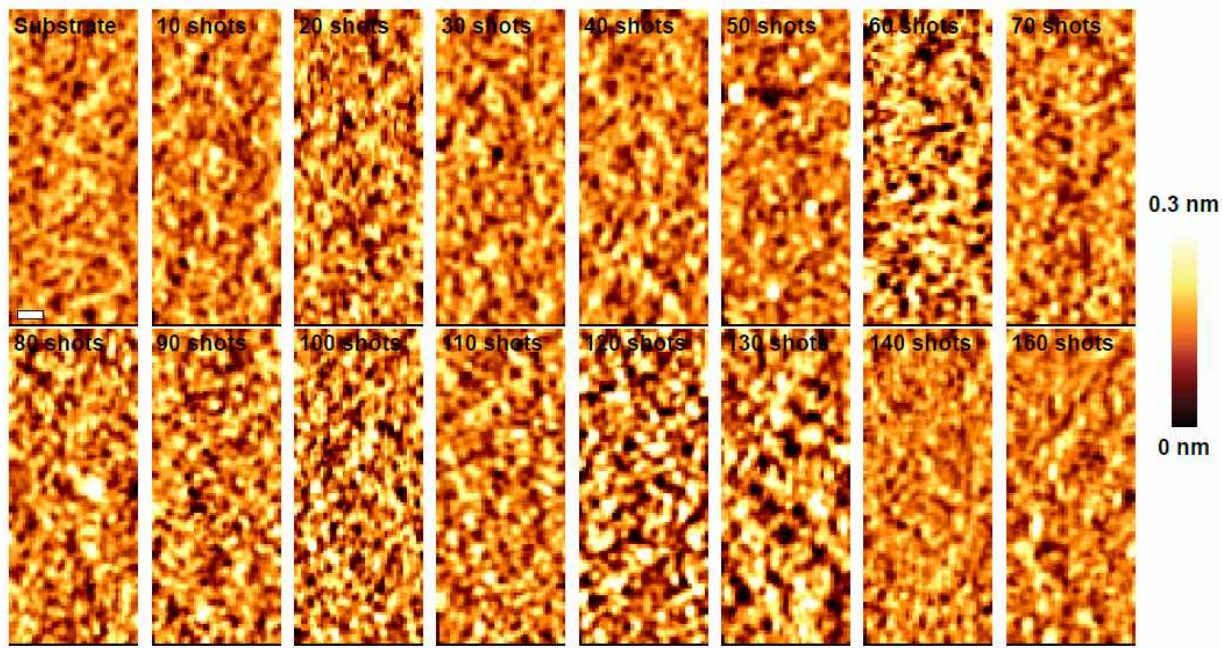

**Fig. S2}.** 1 step images at each shot cropped from red rectangle as indicated in Fig. 2. Scale bar is 40 nm.

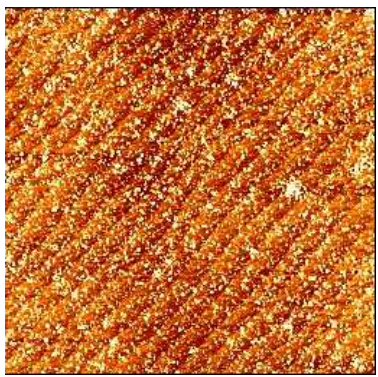

**Fig. S3.** AFM topography of contaminated SRO surface during intermittent AFM measurement for exposure over 2 weeks.